\documentclass[prl,twocolumn,english,superscriptaddress]{revtex4-2}
\usepackage[T1]{fontenc}
\usepackage[latin9]{inputenc}
\usepackage{graphicx}
\usepackage{amsmath}  
\usepackage{xfrac}
\usepackage{txfonts}
\usepackage[mathscr]{euscript}

\makeatletter
\usepackage{babel}
\usepackage{color}
\usepackage[colorlinks = true,
            linkcolor = blue,
            urlcolor  = blue,
            citecolor = blue,
            anchorcolor = blue]{hyperref}

\makeatother
\begin{document}


%
%

\title{Dirac magnons, nodal lines, and nodal plane in elemental gadolinium}
	
	\author{A. Scheie}
	\email{scheieao@ornl.gov}
	\affiliation{Neutron Scattering Division, Oak Ridge National Laboratory, Oak Ridge, TN 37831, USA}
	
	\author{Pontus Laurell}
	\email{laurell@utexas.edu}
	\affiliation{Computational Sciences and Engineering Division, Oak Ridge National Laboratory, Oak Ridge, TN 37831, USA}
	\address{Department of Physics and Astronomy, University of Tennessee, Knoxville, TN 37996, USA.}
	
	\author{P. A. McClarty}
	\affiliation{Max Planck Institute for the Physics of Complex Systems, N\"othnitzer Str. 38, 01187 Dresden, Germany}
    
	\author{G. E. Granroth}
	\affiliation{Neutron Scattering Division, Oak Ridge National Laboratory, Oak Ridge, TN 37831, USA}

	\author{M. B. Stone}
	\affiliation{Neutron Scattering Division, Oak Ridge National Laboratory, Oak Ridge, TN 37831, USA}
	
	\author{R. Moessner}
	\affiliation{Max Planck Institute for the Physics of Complex Systems, N\"othnitzer Str. 38, 01187 Dresden, Germany}

	\author{S. E. Nagler}
	\affiliation{Neutron Scattering Division, Oak Ridge National Laboratory, Oak Ridge, TN 37831, USA}
	\affiliation{Quantum Science Center, Oak Ridge National Laboratory, Tennessee 37831, USA}
	
	\date{\today}

	\begin{abstract}
	    We investigate the magnetic excitations of elemental gadolinium (Gd) using inelastic neutron scattering, showing that Gd is a Dirac magnon material with nodal lines at $K$ and nodal planes at half integer $\ell$. We find an anisotropic intensity winding around the $K$-point Dirac magnon cone, which is interpreted to indicate Berry phase physics. Using linear spin wave theory calculations, we show the nodal lines have non-trivial Berry phases, and topological surface modes.  We also discuss the origin of the nodal plane in terms of a screw-axis symmetry, and introduce a topological invariant characterizing its presence and effect on the scattering intensity. Together, these results indicate a highly nontrivial topology, which is generic to hexagonal close packed ferromagnets. We discuss potential implications for other such systems. 
	\end{abstract}
	
	\maketitle

Topological materials exhibiting quasiparticles with linear band crossings effectively described by the Dirac equation play an important role at the frontier of condensed matter physics \cite{wehling2014dirac,Banerjee_2020}. The electronic structure of Graphene established it as the prototypical example of a fermionic Dirac material \cite{wehling2014dirac, Castro_Neto_2009}. It was subsequently realized that related physics can occur in systems with bosonic quasiparticles including  among others phonons \cite{Li2018}, photons \cite{Khanikaev2013,Lu_2015}, and more recently, magnons \cite{Fransson_2016,owerre2017magnonic,Pershoguba_2018,Malki_2020, Li_2021, McClarty_2021}.  The interesting topological features of magnon bands are often associated with band degeneracies that can be understood as a consequence of symmetries describable by spin-space groups \cite{Brinkman_1967,corticelli2021spin}. 
Magnon band structures can realize analogs of e.g. Chern insulators and topological semimetals \cite{Malki_2020, Li_2021, McClarty_2021} and can host both Dirac \cite{Fransson_2016, owerre2017magnonic, Li_2017} or Weyl magnons \cite{Li_2016, Mook_2016, Su_2017, Su_2017b, Zhang_2020_magnonic}, as well as exhibit extended one-dimensional nodal degeneracies \cite{Li_2017, Mook_2017, Owerre_2017} and triply-degenerate points \cite{Hwang_2020}.  Consequently 
magnetic systems can also exhibit phenomena similar to those found in topological electronic materials, for example a magnon thermal Hall effect arising from gapped bands with topologically non-trivial Chern numbers \cite{Onose_2010, Ideue_2012, Hirschberger_2015b, Chisnell_2015, Chisnell_2016, Laurell_2018}.
In this work we describe a system with a magnon nodal plane degeneracy, thus further extending the fruitful analogy between topological magnets and topological electronic systems \cite{PhysRevB.93.085427, Wu_2018}.

Dirac band crossings have been observed in the layered local-moment magnetic systems CrI$_3$ \cite{Chen_2018} and  CoTiO$_3$ \cite{Yuan_2020,elliot2020visualization}.  
These systems are related to the honeycomb ferromagnet, a simple bipartite lattice that is the prototypical example of a two-dimensional Dirac magnon system.  One strong indicator of non-trivial topology is an anisotropic ``winding'' intensity around the Dirac point, as seen in CoTiO$_3$ \cite{McClarty_2019,shivam2017neutron}. Dirac magnons have also been observed in the three-dimensional antiferromagnet Cu$_3$TeO$_6$ \cite{Yao2018,Bao2018}. 



In this Letter we use inelastic neutron scattering to measure the magnon spectrum of elemental gadolinium (Gd), showing directly that it is a Dirac material. Gd is a highly isotropic ferromagnet with the hexagonal close packed (HCP) structure that forms a simple three-dimensional bipartite lattice. We demonstrate experimentally that the magnon bands in Gd (i) exhibit Dirac nodal lines with a clear anisotropic winding intensity  and non-trivial Berry phase, and (ii) interestingly also show a nodal plane. 
We discuss the protection of the nodal plane by a combination of a screw-axis symmetry and effective time reversal symmetry, and introduce a $\mathbb{Z}_2$ topological invariant to characterize it. 
Our results suggest that the entire class of rare earth HCP ferromagnets is a simple model system for topological magnetism.    

\begin{figure}
	\centering\includegraphics[width=0.48\textwidth]{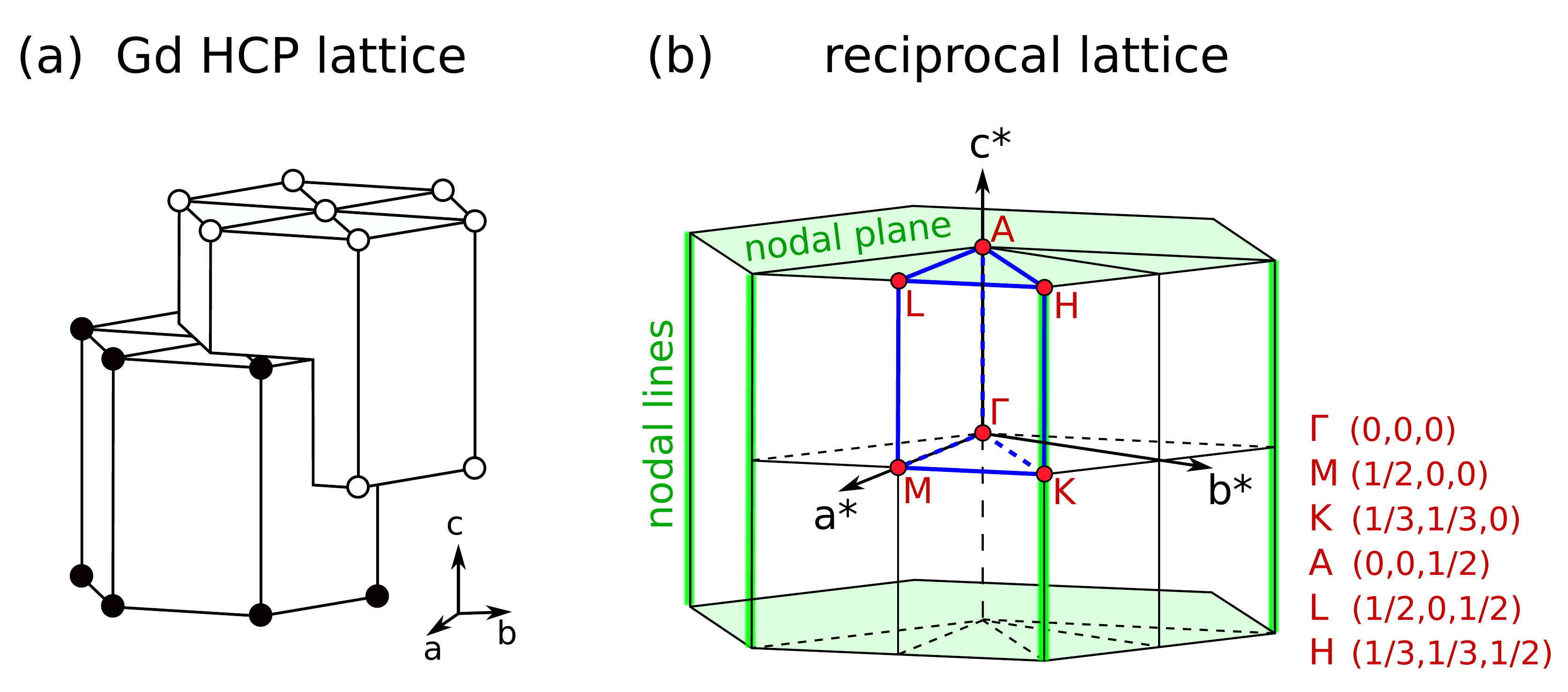}
	\caption{(a) HCP crystal structure of Gd. The lattice is bipartite, with interpenetrating layers of ABAB-stacked triangular lattices. (b) First Brillouin zone of Gd. The dark blue lines delineate the asymmetric unit in reciprocal space, the red dots show the high symmetry points (notated on the right), and the green regions indicate nodal lines at $h=k=1/3$ and nodal planes at $\ell = \pm 1/2$.}
	\label{flo:Gd}
\end{figure}

The Gd HCP structure and its reciprocal lattice are illustrated in Fig. \ref{flo:Gd}.  Gd orders ferromagnetically at $T_c = 293$~K \cite{Nigh_1963, Cable_1968, Urbain_1935}. Although Gd is metallic, the first three valence electrons are completely itinerant and the rest are localized, leaving an effective Gd$^{3+}$ at each site \cite{Moon_1972}. In the half-filled $f$ shell, the orbital angular momentum is effectively quenched leaving $S=7/2$ magnetism \cite{Kip_1953} with near-perfect isotropy and spin-orbit coupling that vanishes to first order. (Small anisotropies do exist in Gd \cite{PhysRevB.79.054406} which influence the direction of the ordered moment \cite{Cable_1968}, but these are of the order 30 $\mu$eV \cite{Franse_1980}---so small that they have never been measured with neutrons.) This makes Gd an ideal material for studying Heisenberg exchange on a hexagonal lattice.

The Gd spin wave spectrum was first measured by Koehler et al. in 1970 \cite{Koehler_1970}; but only along $(hh0)$, $(h\bar{h}0)$, $(h00)$, and $(00\ell)$  directions. These data show a linear magnon band crossing at $K = (1/3,1/3,0)$, indicating a Dirac node and suggesting the possibility of nontrivial topology. The temperature dependence of the Gd magnons was measured in the 1980's \cite{Cable_1985,Cable_1989}, but only along the same symmetry directions as Ref. \cite{Koehler_1970}.
Here we have used SEQUOIA, a modern time of flight spectrometer \cite{Granroth2010,Granroth2006} at the SNS \cite{mason2006spallation}, to measure the Gd inelastic neutron spectrum over the entire Brillouin zone volume.   The sample was a 12~g isotopically enriched $^{160}$Gd single crystal (in fact, the same 99.99\% enriched crystal as was used in Ref. \cite{Koehler_1970}; naturally occurring Gd is highly neutron absorbing) aligned with the $hh\ell$ plane horizontal.  Measurements were carried out at 5 K with incident energies $E_i = 50$~meV and 100~meV. Data were processed with \textit{Mantid} software \cite{Mantid2014}; see the Supplemental Materials \cite{SuppMat} and Ref. \cite{Scheie_Gd_PRB} for further details.  The resulting full data set allows one to directly see topological features in the spectrum.  The data were thoroughly analyzed to determine an accurate spin exchange Hamiltonian: this is discussed in detail in a separate paper \cite{Scheie_Gd_PRB} focusing on the Gd magnetic interactions.  Here we focus on the topological properties of the Gd magnon bands.

\begin{figure}
	\centering\includegraphics[width=0.48\textwidth]{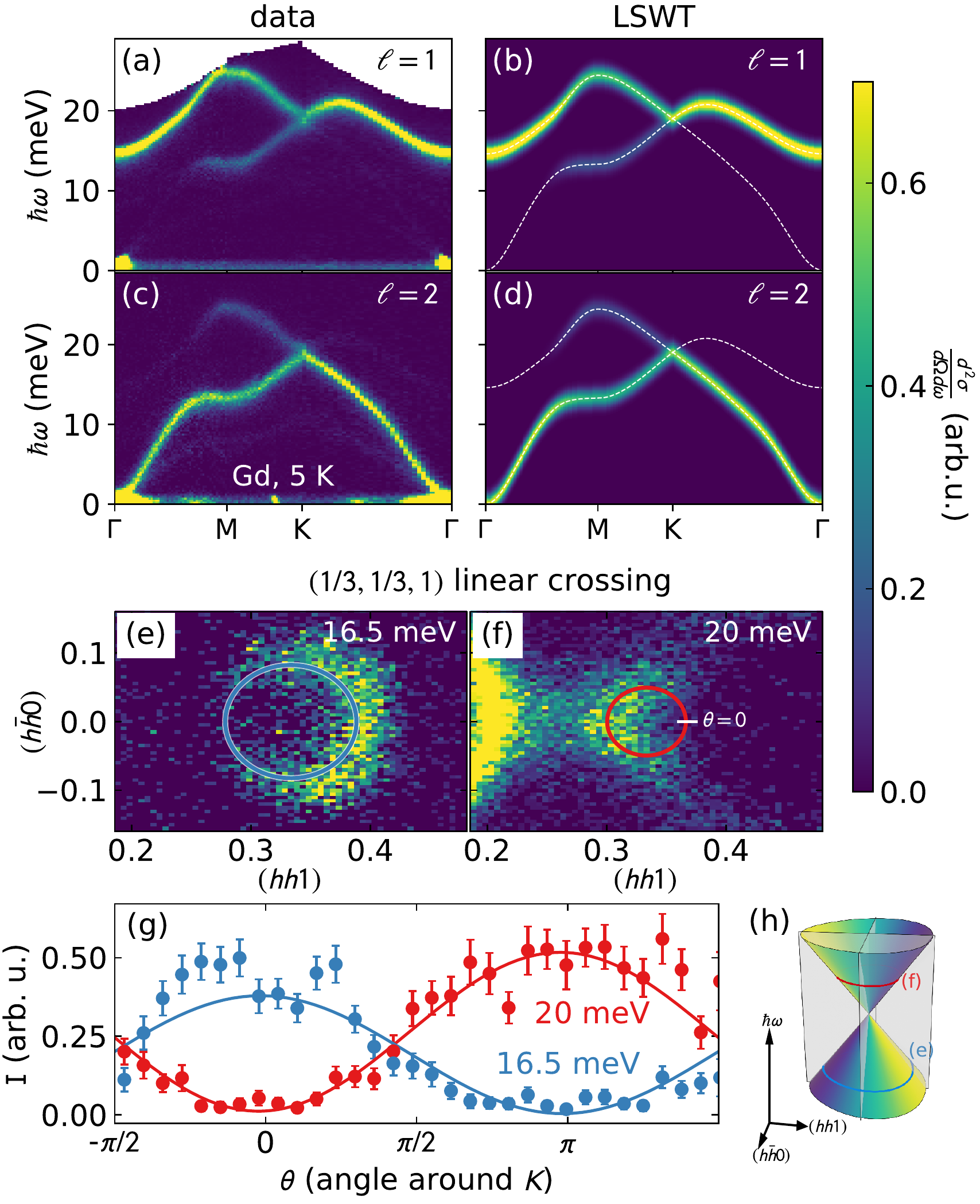}
	\caption{Measured and fitted spin wave spectra of Gd. Panels (a) and (c) show the measured Gd spectra along high-symmetry directions. Panels (b) and (d) show spin wave theory calculated spectra using the best fit Hamiltonian \cite{Scheie_Gd_PRB}. The top row shows the scattering at $\ell=1$, the second row at $\ell=2$. Note the linear band crossing at $K$. 
	Panels (e) and (f) show constant energy slices above and below the band crossing, showing ``intensity arcs''. Panel (g) shows the intensity binned around the circles in (e) and (f), fitted to a sin function. (h) The ``Dirac node'' dispersion surface, with colored circles indicating the slices in panels (e)-(f).}
	\label{flo:SpinWaves}
\end{figure}



Data along high-symmetry directions are shown in Fig. \ref{flo:SpinWaves} alongside the linear spin wave theory (LSWT) fit. 
As this comparison demonstrates, the refined model closely reproduces the measured spectrum. Due to this agreement and the high spin length ($S=7/2$), LSWT is expected to provide 
a good description of Gd.

From a topology perspective, there are two particularly noteworthy features in the Gd scattering: a nodal line degeneracy at $h=k=1/3$ extending along $\ell$, and a nodal plane degeneracy at $\ell=1/2$. We will discuss each in turn.

\begin{figure}
	\centering\includegraphics[width=0.47\textwidth]{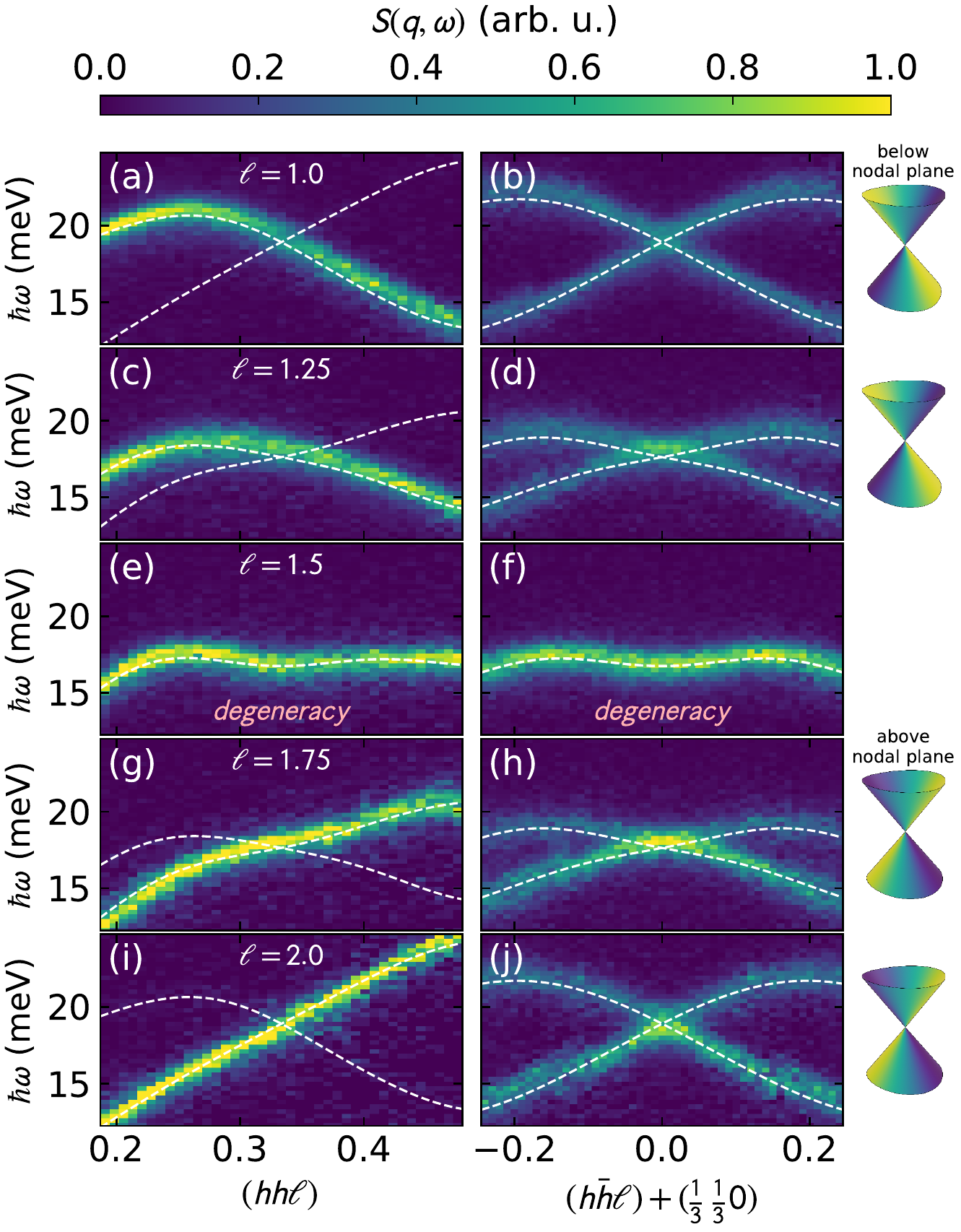}
	\caption{Evolution of the Dirac cone at $K = (\frac{1}{3}\frac{1}{3}\ell)$ as a function of $\ell$. The white dashed lines are calculations using the fitted LSWT Hamiltonian, while the background colormap shows experimental neutron scattering data. The two columns show perpendicular cuts through the $K$ point. As $\ell$ goes from 1 to 2, the cone flattens and inverts, such that the intensity at $\ell=1$ is opposite of $\ell=2$.
	The two LSWT bands are degenerate at $K$ throughout this evolution, yielding a nodal line. 
	Note the emergence of a nodal plane at $\ell = 1.5$, where the two magnon bands degenerate everywhere in the $hk$ plane. To the right are schematics of the Dirac cone, where intensity inverts after crossing the nodal plane.}
	\label{flo:DiracCone}
\end{figure}

The first feature in the data is a linear band crossing at $K$, shown in Fig. \ref{flo:SpinWaves}. As shown in Fig.~\ref{flo:DiracCone}, it extends along $\ell$, making it a nodal line. This band crossing shows an anisotropic intensity pattern [Fig. \ref{flo:SpinWaves} (e)-(h)], where the intensity follows sinusoidal modulation winding around the Dirac cone, inverted above and below the crossing point. A similar intensity winding was seen in CoTiO$_3$ \cite{Yuan_2020, elliot2020visualization}, and is understood to be a signature of the nodal line and nontrivial Berry phase around $(1/3,1/3,\ell)$ \cite{McClarty_2019,shivam2017neutron}. 
(This is similar to a signature of Berry phase physics in graphene seen using polarization-dependent angle-resolved photoemission spectroscopy \cite{Hwang_2011}.) 
Unlike CoTiO$_3$, the offset angle of the intensity winding is zero to within error bars: no anisotropy or off-diagonal exchange shifts the intensity away from the $(hh0)$ line. 


To more firmly establish the topological nature of the nodal line, we turn to linear spin-wave theory \cite{Holstein_1940, Jensen+Mackintosh} and a simplified $J_1-J_2-J_3$ model that qualitatively captures the main features of the full fitted model, including the band crossings,
\begin{equation}
    H   =   J_1 \sum_{\langle i,j\rangle} \mathbf{S}_i \cdot \mathbf{S}_j + J_2 \sum_{\langle\langle i,j\rangle\rangle} \mathbf{S}_i \cdot \mathbf{S}_j + J_3 \sum_{\langle\langle\langle i,j\rangle\rangle\rangle} \mathbf{S}_i \cdot \mathbf{S}_j,  \label{eq:Hsimplified}
\end{equation}
where $J_n$ represents $n$th nearest neighbor exchange. $J_n<0$ 
indicates ferromagnetic 
exchange. (For the values of the exchange couplings, see Ref. \cite{Scheie_Gd_PRB}.) $J_1$ and $J_3$ couple the two sublattices, whereas $J_2$ couple only sites within the same sublattice (within $ab$-planes). 
This model includes three of the four largest magnitude exchange interactions that were determined in the full fit. (Since $J_4$ has a lower coordination number than $J_{1,2,3}$, it only produces a smaller $\ell$-dependent contribution to the energy.)
Details of these calculations are shown in the Supplemental Material \cite{SuppMat}. 

The HCP lattice is inversion symmetric, and the spin-wave Hamiltonian has an effective time-reversal symmetry \cite{Mook_2016, SuppMat}. 
Together, these symmetries guarantee that the Berry curvature vanishes everywhere, and thus HCP Gd does not have non-trivial Chern numbers or Weyl magnons. 
Nevertheless, the same symmetries protect the magnon nodal lines, which are pinned to Brillouin zone corners by threefold rotation symmetry about $\hat{c}$, $C_{3z}$. The topology of the magnon nodal lines can be classified in terms of the Berry phase about a closed contour $\mathcal{C}$,
\begin{equation}
	\gamma_m \left[ \mathcal{C} \right] = \oint_\mathcal{C} \mathrm{d}\mathbf{k} \cdot \mathcal{A}_m \left( \mathbf{k}\right),
\end{equation}
where 
$\mathcal{A}_m	=	i \left\langle u_m \left( \mathbf{k}\right) \middle| \nabla_\mathbf{k} \middle| u_m \left( \mathbf{k}\right) \right\rangle$ is the Berry connection, and $\left| u_m \left( \mathbf{k}\right)\right\rangle \sim \left( \mp \exp(i\phi_{\mathbf{k} }), 1\right)^T$ is the $m$th energy eigenstate of the magnon Hamiltonian. If 
$\mathcal{C}$ is pierced once by a nodal line, it is trivial if $\gamma_m=0$ and non-trivial if $\gamma_m = \pi$. Direct evaluation for Eq.~\eqref{eq:Hsimplified} for Gd shows $\gamma_m\left[ \mathcal{C}\right] = \pm \pi$ for contours surrounding the nodal lines at $K$ and $K'$ \cite{SuppMat}, thus demonstrating their topological nature. It is the nontrivial phase $\phi_\mathbf{k}$ of the wave function $\left| u_m\left(\mathbf{k}\right)\right\rangle$ that generates the Berry phase and the anisotropic intensity, which is proportional to $1\pm \cos \left( \phi_\mathbf{k}\right)$ (plus sign for upper band) and winds about $K$ \cite{SuppMat}.

A second noteworthy feature is a nodal plane. As shown in Fig. \ref{flo:DiracCone}, the Dirac cone flattens and then inverts as $\ell$ increases (plotting between $\ell=1$ and $\ell=2$---the cone at $\ell=0$ is not fully visible due to kinematic constraints of the experiment). In fact, every integer shift in $\ell$ brings an inversion in the Dirac cone intensity, and every half-integer $\ell$ gives a degeneracy in the modes at all $h$ and $k$. This degeneracy, shown in Fig. \ref{flo:DiracCone}(e) and (f) where the Dirac cone is completely flattened, gives rise to a nodal plane. 

Above and below this nodal plane, there is a discontinuous shift in the Dirac cone intensity. This is caused by the phase $\phi_\mathbf{k}$ discontinuously flipping by $\pi$ upon passing through the nodal plane. 
As we discuss in detail in the Supplemental Material \cite{SuppMat}, this nodal plane arises in the HCP ferromagnet from the combination of effective time-reversal and nonsymmorphic twofold screw symmetry $\left\{ C_{2z}, (0,0,1/2) \right\}$, connecting the two sublattices. Spin orientation plays no role in the Heisenberg limit. Any magnetic Hamiltonian which maintains these symmetries will also have a symmetry-protected nodal plane.

We can describe the nodal plane more formally by defining 
a $\mathbb{Z}_2$ topological invariant, which 
changes discontinuously across the nodal plane. Such an invariant can either be defined in terms of the Pfaffian of a transformed magnon Hamiltonian \cite{SuppMat}, or in terms of wavefunction properties. Here we focus on the latter. 
We define $\nu^m_{\mathbf{k}}\equiv {\rm sgn}\ \langle u_m \left(\mathbf{k}\right) \vert \sigma_1 \vert u_m\left(\mathbf{k}\right) \rangle$, 
where $\sigma_1$ is the first Pauli spin matrix. 
If we choose a reference wavevector $\mathbf{k}$ and $\mathbf{k}'\equiv \mathbf{k} + (0,0,\delta k_z)$ the difference $1/2\vert \nu_{\mathbf{k}}-\nu_{\mathbf{k}'}\vert$ counts the number of times the nodal plane is crossed (and thus the number of times the intensity inverts) modulo two. 


\begin{figure}
	\centering\includegraphics[width=0.48\textwidth]{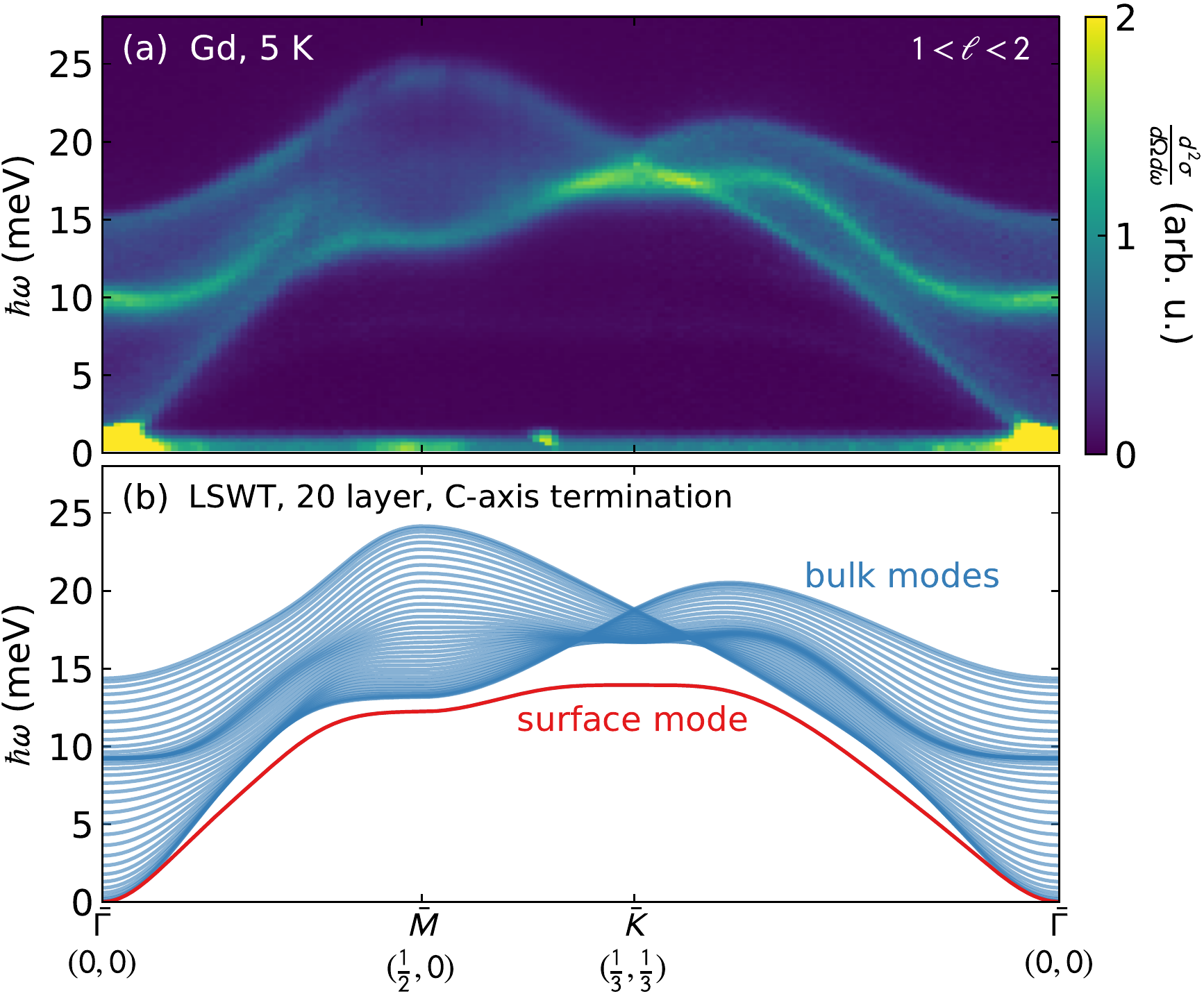}
	\caption{Surface magnons in Gd. (a) In-plane high-symmetry cuts of 5~K Gd scattering integrated from $\ell=1$ to $\ell=2$. (b) Linear spin wave theory (LSWT) calculated modes for a 20-layer Gd slab using the best fit Hamiltonian. Note that, because of the finite extent along $\hat{c}$, $\ell$ is no longer a good quantum number, and the magnon modes from each layer form a continuum between $\ell=1$ and $\ell=2$, such that the magnon modes strongly resemble the integrated data in panel (a). The $c$-axis termination surface magnon mode, shown in red, lies outside this continuum at lower energies, and is thus distinct from bulk magnons. 
	}
	\label{flo:BulkSurface}
\end{figure}

%
Although the nodal plane is not expected to produce a topological surface state \cite{Wu_2018, Xiao_2020}, the nodal lines are.
To investigate this, we theoretically considered the simplest geometry for surface modes: a slab of a finite number of triangular lattice layers along $\hat{c}$ as shown in Fig.~\ref{flo:Gd}. This was done for the full fitted LSWT model (26 neighbor exchange terms) using the \textit{SpinW} software \cite{SpinW} by creating a supercell geometry with and without periodic boundary conditions in the $c$ direction (the $c$ termination was generated by creating a blank space at the top of the physical layers, effectively breaking periodicity). The result is shown in Fig.~\ref{flo:BulkSurface} for $20$ Gd unit cells ($40$ triangular lattice layers). LSWT [Fig.~\ref{flo:BulkSurface}(b)] shows the presence of a clear surface mode, emerging from the bulk modes projected into the 2D surface Brillouin zone. Since inelastic neutron scattering is not a surface probe we cannot resolve the same mode in the data, but nevertheless find qualitative agreement with the bulk modes [Fig.~\ref{flo:BulkSurface}(a)].

It should be emphasized that neither of these degeneracies---the nodal line at $h=k=1/3$ and the nodal plane at $\ell=1/2$---depend sensitively upon the details of the magnetic exchange Hamiltonian. 
On the HCP lattice, they appear with both the simplest nearest neighbor ferromagnetic exchange interaction, or with any number of further neighbor exchanges---so long as they are all Heisenberg exchanges and the ground state remains ferromagnetic,  preserving effective time-reversal symmetry (this was first noted by Brinkman in 1967 \cite{Brinkman_1967} and the topological consequences have been explored in Ref. \cite{corticelli2021spin}). Thus, although the further neighbor exchange interactions are important for understanding the wiggles in Gd's magnon dispersion, they are not important for understanding the topology.

These experiments and calculations were carried out on Gd, which has near-perfect isotropic Heisenberg exchange. However, because of the intrinsic connection between symmetry, degeneracy, and topology \cite{Cracknell_1970,Brinkman_1967,Narang_2021,Watanabe_2018,corticelli2021spin} similar topological features can be 
expected in more anisotropic ferromagnetic HCP metals such as Tb \cite{Lindgard_1978,Moller_1968}, Dy \cite{Lindgard_1978,Nicklow_1971}, and hexagonal Co \cite{perring1995high}. 
(However, for Co one must consider the effects of itinerancy and continuum scattering likely eliminate the observability of Dirac magnons in HCP Co \cite{Do_2021,Okumura_2019,Skovhus_2021}.)

From a topological magnon perspective, it is particularly interesting to consider the addition of interactions breaking the symmetries protecting the nodal degeneracies. One choice which can break the effective time-reversal symmetry is the 
Dzyaloshinskii-Moriya (DM) exchange interaction \cite{Moriya_1960} 
\begin{equation}
    H = \sum_{ij} \bf{D} \cdot ( {\bf S}_i \times {\bf S}_j),
\end{equation}
where $\bf{D}$ is the DM vector. 
Like on the honeycomb lattice \cite{McClarty_2021}, it is symmetry-allowed on the HCP lattice second nearest neighbor bonds. 


It is easily shown on the level of LSWT that easy axis or easy plane single-ion anisotropy preserves the extended degeneracies as the effective time reversal symmetry, originating from spin-space symmetries, is preserved, whereas DM exchange with out-of-plane $\mathbf{D}$ vector lifts the $K$-point and nodal plane degeneracy while leaving a grid of $\ell=1/2$ nodal lines, giving rise to potential chiral surface magnon modes \cite{Scheie_Gd_PRB}. However, the true situation is more complicated for anisotropic rare earth HCP ferromagnets such as Tb or Dy. In such cases, the strong spin-orbit coupling may induce other symmetry-allowed off-diagonal exchange, which would in turn affect the surface modes. This means that that inducing chiral surface modes in these materials may prove a challenge. Full characterization of other HCP ferromagnets spin exchange Hamiltonian is necessary to determine the possibility of directional surface modes.

In conclusion, we have shown that the magnetic excitation spectrum of elemental gadolinium contains nodal line and nodal plane degeneracies, which are directly visible in the experimental data. The nodal line around $K$ shows anisotropic intensity characteristic of nontrivial topology, and Berry phase calculations confirm this to be so. We also identify a nodal plane in the data, derive the symmetry requirements for such a feature, and propose an invariant describing its topology.
These results have implications not just for Gd, but for all HCP ferromagnets, as the topological features are generic to the lattice. 
Other consequences of the HCP topology may exist---particularly concerning the nodal plane---but these are left for future study.

\section*{Acknowledgments}

We acknowledge helpful discussions with Satoshi Okamoto.
This research used resources at the Spallation Neutron Source, a DOE Office of Science User Facility operated by the Oak Ridge National Laboratory. The research by P.L. was supported by the Scientific Discovery through Advanced Computing (SciDAC) program funded by the US Department of Energy, Office of Science, Advanced Scientific Computing Research and Basic Energy Sciences, Division of Materials Sciences and Engineering. The work by SEN is supported by the Quantum Science Center (QSC), a National Quantum Information Science Research Center of the U.S. Department of Energy (DOE).

%

\newpage

\quad

\newpage

\def\theequation{S\arabic{equation}}
\def\thefigure{S\arabic{figure}}
\setcounter{equation}{0}
\setcounter{figure}{0}

\section*{Supplemental Information for Dirac magnons, nodal lines, and nodal plane in elemental gadolinium}

This supplement contains I. parameters for the experiment, II. a discussion of symmetry properties and topological invariants for the general hexagonal closed packed (HCP) ferromagnet spin-wave problem, and III. an explicit linear spin-wave theory (LSWT) treatment of the spectrum and topology of the $J_1-J_2-J_3-J_4$ model.

\section{Experiment parameters}

For the SEQUOIA measurement, we ran the $T0$ chopper at 90~Hz, Fermi 1 chopper at 120~Hz, Fermi 2 chopper at 360~Hz for $E_i = 50$~meV. For  $E_i = 100$~meV we ran the same configuration but Fermi 2 chopper at 540~Hz. The sample was rotated in one degree steps to measure the inelastic spectra, and the data were reduced and symmetrized \cite{Mantid2014} to fill out the full Brillouin zone.


\section{Symmetry properties and topology of the nodal plane} 
\label{sec:symtop}

The nodal plane lives on the hexagonal boundaries of the Brillouin zone. Here we show that it is enforced by effective time reversal and nonsymmorphic symmetries.

Gadolinium crystallizes into a HCP structure with space group \#$194$ or P$6_3$/mmc. We place an origin midway between triangular layers on a line extending perpendicular to the triangular planes at the centroid of one of the triangles. This group has $24$ generators besides translations. Some of the nontrivial elements of this group are as follows:
\begin{enumerate}
	\item Threefold rotation about $\hat{z}$: $C_{3z}$ and $C_{3z}^2$,
	\item A screw composed of $C_{2z}$ and a translation along $\hat{z}$ through $(0,0,1/2)$,
	\item Twofold rotation axes along $(1,0)$, $(0,1)$ and $(1,1)$ through the origin,
	\item Twofold rotation axes in-plane $30$ degrees rotated about $\hat{z}$ from those above followed by a $(0,0,1/2)$ translation,
	\item Sixfold screw axis with $(0,0,1/2)$ translation,
	\item Inversion about the origin,
\end{enumerate}
and compositions of these.

For our purposes, an important observation is that the group is nonsymmorphic with a twofold screw axis that we denote $\left\{ C_{2z}, (0,0,1/2) \right\}$. There is also a glide symmetry that can be obtained by composing the screw and the inversion symmetries. If a general lattice position is denoted $m\mathbf{a}_1 + n\mathbf{a}_2 +l\mathbf{a}_3$ and a general wavevector by $k_1\mathbf{b}_1 + k_2\mathbf{b}_2 +k_3\mathbf{b}_3$, the screw symmetry acts on the sites as $(m,n,l)\rightarrow (-m,-n,l+1/2)$ and the sublattice label swaps. Thus applying the screw twice is equivalent to a translation through one primitive vector out of plane. It follows that the action of the screw on a magnon state is
\begin{align}
& U(\left\{ C_{2z}, (0,0,1/2) \right\})\vert k_1,k_2,k_3; 1 \rangle \rightarrow \vert -k_1,-k_2,k_3; 2 \rangle \\
& U(\left\{ C_{2z}, (0,0,1/2) \right\})\vert k_1,k_2,k_3; 2 \rangle \rightarrow e^{2\pi i k_3} \vert -k_1,-k_2,k_3; 1 \rangle,
\end{align}
where we take a (periodic) Fourier transform convention with $H(\mathbf{k})= H(\mathbf{k}+\mathbf{G})$.

Importantly, the magnon Hamiltonian also satisfies an effective time reversal symmetry. 
Physical time reversal is broken by the ferromagnetic order, but the fact that the magnetic Hamiltonian has only rotationally invariant couplings tells us that the system is left invariant under the application of time reversal followed by a rotation of the moments through axes perpendicular to the moments, which can easily be verified using the notation of Ref.~\cite{Mook_2016}. This spin-space symmetry is anti-unitary and therefore acts like an effective time reversal symmetry $\mathcal{T}^*$. It is inherited by the magnon Hamiltonian where it acts as $\mathbf{k}\rightarrow -\mathbf{k}$ and complex conjugation. 

The degeneracy on the hexagonal Brillouin zone boundary is enforced by the product of time reversal and the screw symmetries: $\mathcal{T}^* U(\left\{ C_{2z}, (0,0,1/2) \right\})$. In particular, the square of this symmetry element is $\exp( - 2\pi i k_3) = -1$ on the upper and lower Brillouin zone faces where $k_3=\pm 1/2$ implying that there is a Kramers degeneracy in the two-band magnon model on this surface. 

We may look at this from the perspective of a general two-band Hamiltonian
\begin{align}
H(\mathbf{k}) = \left( \begin{array}{cc} A(\mathbf{k}) & B(\mathbf{k}) \\ B^*(\mathbf{k}) & A'(\mathbf{k})  \end{array}\right).
\end{align}
Time reversal symmetry forces $A(\mathbf{k})$ and $A'(\mathbf{k})$ to be even in momentum and $B(\mathbf{k}) = B^*(-\mathbf{k})$. It is now convenient to switch from $k_1, k_2, k_3$ to Cartesian $k_x,k_y,k_z$.
The twofold screw symmetry acts as $G(k_z)H(-k_x,-k_y,k_z)G^\dagger(k_z)  = H(k_x,k_y,k_z)$ or
\begin{align}
{\small
	\left(  \begin{array}{cc} 0 & 1 \\ e^{-ic k_z} & 0  \end{array} \right)\left(  \begin{array}{cc} A(-k_x,-k_y,k_z) & B(-k_x,-k_y,k_z) \\ B^*(-k_x,-k_y,k_z) & A'(-k_x,-k_y,k_z)  \end{array} \right)\left(  \begin{array}{cc} 0 & e^{ic k_z} \\ 1 & 0  \end{array} \right)} \nonumber \\
= H(k_x,k_y,k_z)
\end{align}
where $G^2(k_z)=\exp(-i c k_z)$, which implies that
\begin{align}
A'(-k_x,-k_y,k_z) = A(k_x,k_y,k_z).
\label{eq:2foldscrewA}
\end{align}
On the zone boundary $k_z = \pm\pi$, the constraint from time reversal that $A$ and $A'$ are even in momentum now implies that $A'(k_x,k_y)=A(k_x,k_y)$. Now consider $B(\mathbf{k})$. The screw symmetry implies that 
\begin{align}
e^{ic  k_z} B^*(-k_x,-k_y,k_z) = B(k_x,k_y,k_z)
\label{eq:2foldscrewB}
\end{align}
and, with time reversal $B(\mathbf{k}) = B^*(-\mathbf{k})$ at $k_z=\pm \pi$ we find that $B$ must vanish. We have therefore directly shown the presence of the nodal plane in the two band model.

It is worth pointing out that the symmetry is high enough to force $B(k_x,k_y,\pm \pi)$ to vanish whether time reversal is present or not. We consider only the two-fold screw symmetry and inversion. Inversion has the effect of taking $\mathbf{k}$ to $-\mathbf{k}$ and swapping the sublattices so $B(k_x,k_y,k_z)= B^*(-k_x,-k_y,-k_z)$. Recalling the constraint from the screw symmetry $e^{i c k_z} B^*(-k_x,-k_y,k_z) = B(k_x,k_y,k_z)$ we obtain, at $k_z=\pm \pi$, that $B$ must vanish. However in this case $A$ is not constrained to equal $A'$ through these symmetries alone so the time reversal symmetry is essential to the nodal plane in this system.

Since there is both inversion and time reversal in the Heisenberg model, the topological charge of the nodal planes is zero as it is for the nodal lines. Another way of putting this is that there are no sources of Berry flux (as it is zero by symmetry).  

\begin{figure}
	\centering\includegraphics[width=0.49\textwidth]{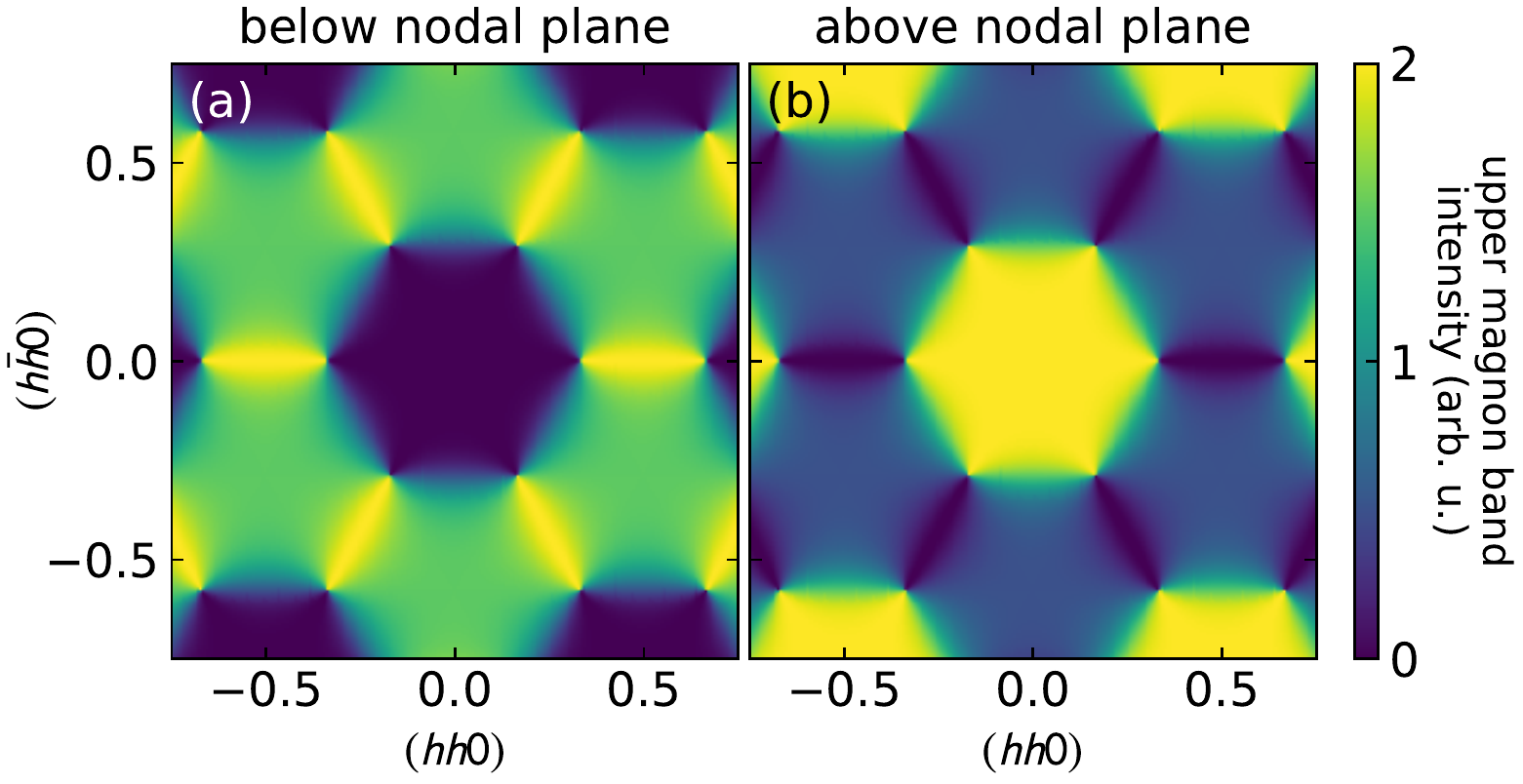}
	\caption{Calculated intensity of the upper (higher energy) magnon mode across the Brillouin zone below (a) and above (b) the $\ell=1/2$ nodal plane for the HCP ferromagnet. When the nodal plane is passed, the intensity pattern inverts.}
	\label{flo:NodalPlaneIntensity}
\end{figure}

Although the topological charge of the nodal planes is zero, one may ask whether there is an invariant for the nodal planes analogous to the $\pi$ winding of the Berry phase around nodal lines. We focus on the nontrivial phase of the wavefunction $\phi_{\mathbf{k}}$ in an eigenstate of the Hamiltonian at $\mathbf{k}$: 
\begin{align}
\psi_{\mathbf{k}-} = \left( \begin{array}{c} -\exp(i\phi_{\mathbf{k} }) \\ 1 \end{array} \right) \hspace{1cm} \psi_{\mathbf{k}+} = \left( \begin{array}{c} \exp(i\phi_{\mathbf{k} }) \\ 1 \end{array} \right) \label{eq:general:psi}
\end{align} 
This phase has observable consequences as the intensity in each band is proportional to $1\pm \cos(\phi_\mathbf{k})$ (plus sign for upper band). Around the nodal lines, the phase winds and this is responsible for the highly anisotropic intensity in their vicinity. Everywhere inside the zone the phase is completely smooth. However, the presence of the nodal plane has an unmistakable effect on the phase: it flips by $\pi$ discontinuously on passing through the nodal plane in the $k_z$ direction. This results in a discontinuous change in the intensities when passing through the nodal plane, as shown in Fig. \ref{flo:NodalPlaneIntensity} (also see Fig. 3 in the main text). The appropriate topological invariant picks up this phase flip. 

How can we see it at the level of the Hamiltonian?  Now take, for convenience, the Fourier transform convention including basis vectors (i.e. $\sum_{i} F_{ia} \exp(i\mathbf{k}\cdot (\mathbf{R}_i + \mathbf{r}_a))$ where $\mathbf{r}_a$ is a basis vector and $\mathbf{R}_i$ a primitive lattice vector). The phase $\phi_{\mathbf{k}}$ originates from the off-diagonal components and these have a $k_z$ dependence that looks like $\cos(k_z)$. Within the zone, this merely modulates the size of the off-diagonal components without changing the phase. This further implies that the winding of the Berry phase within the zone along $k_z$ is trivial. However on passing through the zone boundary along $k_z$, $\cos(k_z)$ changes sign which is equivalent to $\phi_{\mathbf{k}} \rightarrow \phi_{\mathbf{k}} + \pi$. One way of characterizing this phase change is to remove the diagonal components of the Hamiltonian as they merely shift the bands. This done, the Hamiltonian is 
$H(\mathbf{k}) = f_1(\mathbf{k})\sigma_1 + f_2(\mathbf{k})\sigma_2$ and a unitary transformation brings this into the form
\begin{align}
\tilde{H}(\mathbf{k}) = \left( \begin{array}{cc} 0 & -iq \\ iq & 0  \end{array}\right).
\end{align}
Define $\mu_{\mathbf{k}}\equiv {\rm sgn}\ \mathrm{Im}\left[{\rm pf} \tilde{H}(\mathbf{k})\right]$ where ${\rm pf}$ denotes the Pfaffian. This number is smooth in the zone and changes discontinuously across the nodal plane. Thus, if we choose a reference wavevector $\mathbf{k}$ and $\mathbf{k}'\equiv \mathbf{k} + (0,0,\delta k_z)$ the difference $1/2\vert \mu_{\mathbf{k}}-\mu_{\mathbf{k}'}\vert$ counts the number of times the nodal plane is crossed modulo two. If the protecting time reversal and screw symmetry is broken, the phase will tend to vary smoothly along suitably chosen paths in momentum space. If the symmetry is in place, the invariant is a robust diagnostic of the presence of the nodal plane regardless of the nature of the magnetic interactions. Another way of formulating an invariant for this system is through the quantity $\nu_{\mathbf{k}}\equiv {\rm sgn}\ \langle \psi_{\mathbf{k}\pm} \vert \sigma_1 \vert \psi_{\mathbf{k}\pm} \rangle$ and associated invariant $1/2\vert \nu_{\mathbf{k}}-\nu_{\mathbf{k}'}\vert$ as the matrix element is essentially $\cos \phi_{\mathbf{k} }$. This wave-function-based invariant is also shown in the main text.

\section{Spin-wave theory of the \texorpdfstring{$J_1-J_2-J_3-J_4$}{J1-J2-J3-J4} model}

\begin{figure}
	\centering\includegraphics[width=0.9\columnwidth]{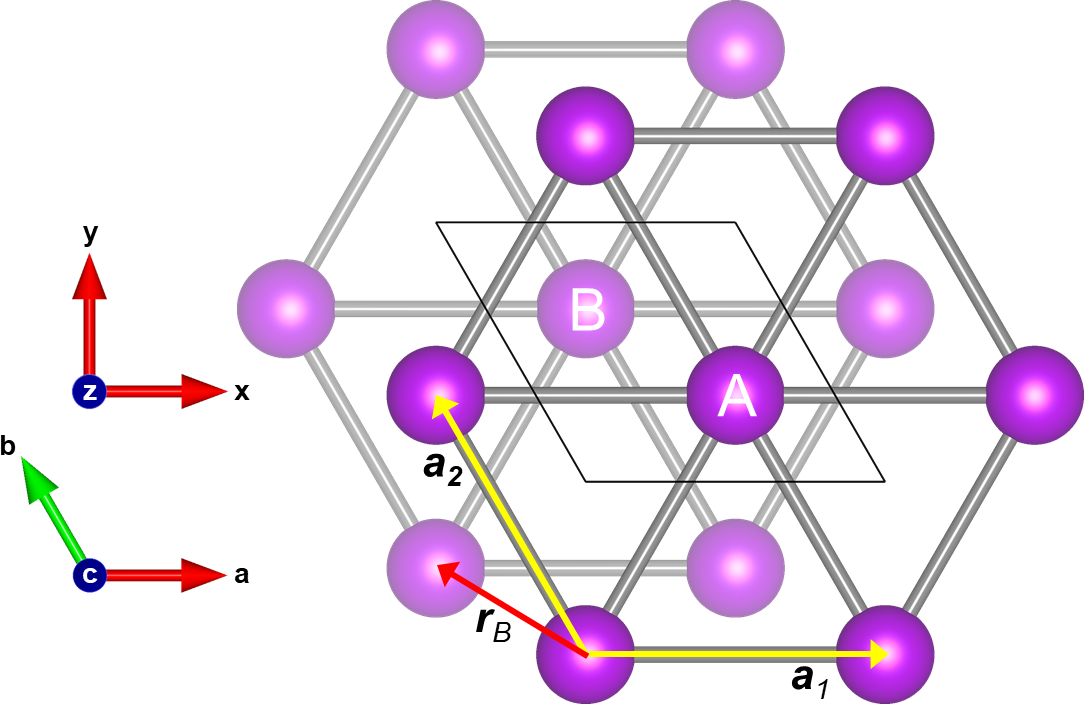}
	\caption{HCP crystal structure of Gd, with layers of triangular lattices. The two sublattices are labeled $A,B$. Both the crystallographic $(abc)$ and Cartesian $(xyz)$ coordinate systems are indicated, along with in-plane lattice vectors $\mathbf{a}_{1/2}$ and basis vector $\mathbf{r}_B$. The vertical lattice vector, $\mathbf{a}_3$, points out of the depicted plane.}
	\label{flo:Gd:supp}
\end{figure}
Here we provide analytical LSWT results for the simplified model. Although we limited the discussion in the main text to a $J_1-J_2-J_3$ model, it is straightforward to include also $J_4$ in the explicit LSWT calculations, and we will do so here by considering
\begin{equation}
H   =   \sum_{n=1}^4 \sum_{i,j} \, J_n^{i,j} \mathbf{S}_i \cdot \mathbf{S}_j,  \label{eq:HsimplifiedS}
\end{equation}
where $J_n^{i,j}=J_n$ if sites $i,j$ are $n$th nearest neighbors, and $J_n^{i,j}=0$ otherwise. Similarly to Ref.~\cite{Jensen+Mackintosh}, we describe Gd as a two-sublattice ferromagnet consisting of ABAB-stacked triangular lattice layers, as shown in Fig.~\ref{flo:Gd:supp}. Denoting the in- and out-of-plane lattice constants by $a$ and $c$, respectively, the lattice ($\mathbf{a}_i$, $i=1,2,3$) and basis vectors ($\mathbf{r_{A/B}}$) can be chosen (expressed in the $(xyz)$ coordinate system indicated in Fig.~\ref{flo:Gd:supp})
\begin{equation}
\mathbf{a}_1	=	\left( a, 0, 0\right), \quad \mathbf{a}_2	=	\left( -\frac{a}{2}, \frac{a \sqrt{3}}{2}, 0\right), \quad \mathbf{a_3}	=	\left( 0,0,c\right),
\end{equation}
\begin{equation}
\mathbf{r}_A	=	\left( 0,0,0\right),\quad \mathbf{r}_B	=	\left( \frac{-a}{2}, \frac{a}{2\sqrt{3}}, \frac{c}{2} \right).
\end{equation}
The resulting reciprocal lattice vectors are
\begin{align}
\mathbf{b}_1	&=	\frac{2\pi}{a} \left( 1, \frac{1}{\sqrt{3}}, 0\right),\quad \mathbf{b}_2	=	\frac{4\pi}{a\sqrt{3}} \left( 0, 1, 0\right),\\
\mathbf{b}_3	&=	\frac{2\pi}{c} \left( 0, 0, 1\right).
\end{align}

To lowest order in the Holstein-Primakoff expansion,
\begin{equation}
S_i^+	=	\sqrt{2S} a_i,\quad S_i^- = \sqrt{2S}a_i^\dagger,\quad S_i^z = S-a_i^\dagger a_i,
\end{equation}
for $i\in A$, and with $a_i\rightarrow b_i$ for $i\in B$. After substitution into Eq.~\eqref{eq:HsimplifiedS}, keeping terms quadratic in creation and annihilation operators, and Fourier transforming we obtain the LSWT Hamiltonian
\begin{align}
H^{(2)}	&=	\sum_\mathbf{k} \left\{ C_\mathrm{d} \left( \mathbf{k}\right) \left( a_\mathbf{k}^\dagger a_\mathbf{k} + b_\mathbf{k}^\dagger b_\mathbf{k} \right) \right.\nonumber\\
&\left. + C_\mathrm{o} \left( \mathbf{k}\right) a_\mathbf{k}^\dagger b_\mathbf{k}	+ C_\mathrm{o} \left( -\mathbf{k}\right) b_\mathbf{k}^\dagger a_\mathbf{k} \right\},
\end{align}
where
\begin{align}
C_\mathrm{d} \left( \mathbf{k}\right)	&=	12J_2 S \left( \gamma_2 \left( \mathbf{k}\right) -1 \right) + 4J_4S \left( \gamma_4 \left( \mathbf{k}\right)-1\right)	\nonumber\\
&- 12J_1S -12J_3S ,	\\
C_\mathrm{o} \left( \mathbf{k}\right)	&=	12S \left[ J_1 \gamma_1 \left(\mathbf{k}\right) + J_3 \gamma_3 \left(\mathbf{k}\right) \right],
\end{align}
$\gamma_n \left( \mathbf{k}\right)	=	\frac{1}{z_n}\sum_{\vec{\delta}_n} e^{-i \mathbf{k} \cdot \vec{\delta}_n}$, $z_n$ is the number of $n$th nearest neighbors, and $\vec{\delta}_n$ are the $n$th nearest neighbor vectors. Explicitly, the neighbor vectors are given by
\begin{align}
\vec{\delta}_1	&\in	\left\{ \mathbf{r}_B^\eta, \mathbf{a}_1 + \mathbf{r}_B^\eta, - \mathbf{a}_2 + \mathbf{r}_B^\eta,\right.\nonumber\\
&\left. \mathbf{r}_B^\eta - \mathbf{a}_3, \mathbf{a}_1 + \mathbf{r}_B^\eta - \mathbf{a}_3,  - \mathbf{a}_2 + \mathbf{r}_B^\eta -\mathbf{a}_3 \right\},\\
\vec{\delta}_2	&\in	\pm\left\{ \mathbf{a}_1, \mathbf{a}_2, \mathbf{a}_1+\mathbf{a}_2\right\},\\
\vec{\delta}_3	&\in	\left\{ \mathbf{a}_1 + \mathbf{a}_2 + \mathbf{r}_B^\eta, \mathbf{a}_1 - \mathbf{a}_2 + \mathbf{r}_B^\eta, - \mathbf{a}_1 - \mathbf{a}_2 + \mathbf{r}_B^\eta,\right.\\
&\left. \mathbf{a}_1 + \mathbf{a}_2 + \mathbf{r}_B^\eta - \mathbf{a}_3, \mathbf{a}_1 - \mathbf{a}_2 + \mathbf{r}_B^\eta - \mathbf{a}_3, - \mathbf{a}_1 - \mathbf{a}_2 + \mathbf{r}_B^\eta - \mathbf{a}_3 \right\},\nonumber\\
\vec{\delta}_4	&\in	\pm \mathbf{a}_3,
\end{align}
where $\mathbf{r}_B^\eta=\eta\mathbf{R}$, and $\eta=1$ ($\eta=0$) for the Fourier convention including basis vectors (the periodic Fourier convention). Note that HCP lattice sites are not centers of inversion, and that the $\vec{\delta}_{1/3}$ vectors connect sublattices in the direction $A\rightarrow B$. (For $B\rightarrow A$, simply use $\vec{\delta}'_{1/3}=-\vec{\delta}_{1/3}.$) With these vectors we obtain
\begin{align}
\gamma_2 \left( \mathbf{k}\right)	&=	\frac{1}{3} \left[\cos (a  k_x)+2 \cos \left(\frac{a  k_x}{2}\right) \cos \left(\frac{\sqrt{3} a  k_y}{2} \right)\right],	\label{eq:gamma2}	\\
\gamma_4 \left( \mathbf{k}\right)	&=	\cos \left( c k_z \right),	\label{eq:gamma4}
\end{align}
both of which are manifestly real-valued, and for $\eta=1$
\begin{align}
\gamma_1^{\eta=1}	&=\frac{1}{6} \left(1+e^{i c k_z}\right) \left(e^{\frac{i a  k_x}{2}}+e^{\frac{1}{2} i a  \left(2 k_x+\sqrt{3}
	k_y\right)}+e^{\frac{1}{2} i \sqrt{3} a  k_y}\right) \nonumber\\
&\times e^{-\frac{1}{6} i \left(3 c k_z+3 a  k_x+2 \sqrt{3} a  k_y\right)},	\label{eq:gamma1}	\\
\gamma_3^{\eta=1}	&=	\frac{1}{6} \left(1+e^{i c k_z}\right) \left(e^{2 i a  k_x}+e^{i a  \left( k_x+\sqrt{3} k_y\right)}+1\right)	\nonumber\\
&\times e^{-\frac{1}{6} i \left(3 c k_z+6 a  k_x+2 \sqrt{3} a  k_y\right)}.	\label{eq:gamma3}
\end{align}
which are generally complex-valued. In the periodic Fourier convention we instead find
\begin{align}
\gamma_1^{\eta=0}	&=\frac{1}{6} \left(1+e^{i c k_z}\right) \left(1 + e^{-iak_x}+e^{\frac{1}{2} i a  \left( \sqrt{3} k_y - k_x\right)}  \right),	\label{eq:gamma1periodic}	\\
\gamma_3^{\eta=0}	&=	\frac{1}{6} \left(1+e^{i c k_z}\right) \left(e^{i a  k_x}+e^{i a \sqrt{3} k_y}+e^{ia\left( 2k_x + \sqrt{3}k_y \right)}\right)	\nonumber\\
&\times e^{-\frac{1}{2} ia \left(3 k_x + \sqrt{3} k_y\right)}.	\label{eq:gamma3periodic}
\end{align}
These functions all satisfy $\gamma_n \left( -\mathbf{k}\right) = \gamma_n^\star \left( \mathbf{k}\right)$, where $^\star$ denotes complex conjugate. While $\gamma_{2,4}\left( \mathbf{k}\right)$ are invariant under both $C_3$ and $C_6$ rotations about $\hat{k_z}$, $\gamma_{1,3}\left(\mathbf{k}\right)$ [and thus also $h\left(\mathbf{k}\right)$] has $C_3$ symmetry but not $C_6$.

Since there are no anomalous terms in the magnon Hamiltonian $H^{(2)}$, it can be diagonalized unitarily. We write
\begin{align}
H^{(2)}	&=		\text{const} + \sum_{\mathbf{k}} \mathbf{X}^\dagger_\mathbf{k} h\left(\mathbf{k}\right) \mathbf{X}_\mathbf{k},
\end{align}
where
\begin{align}
\mathbf{X}_\mathbf{k}	&=	\left( a_\mathbf{k}, b_\mathbf{k}	\right)^T,	\quad	h\left( \mathbf{k}\right)	=	\left( \begin{array}{cc} C_\mathrm{d} \left( \mathbf{k}\right)	& C_\mathrm{o} \left( \mathbf{k}\right)	\\
C_\mathrm{o}^\star \left( \mathbf{k}\right)	&	C_\mathrm{d} \left( \mathbf{k}\right)	\end{array}\right),	\label{eq:matrixHamiltonian}
\end{align}
and diagonalize $h(\mathbf{k})$. (We note that while the form of operators such as $h(\mathbf{k})$ depends on the Fourier convention, observables do not.) This yields eigenvalues
\begin{align}
\epsilon_{1,2}	&=	-2S \left( 6J_1 + 6J_2 + 6J_3 + 2 J_4 \right) + 12 J_2 S \gamma_2 \left( \mathbf{k}\right)	\nonumber\\	
&+ 4 J_4 S \gamma_4 \left( \mathbf{k} \right) \mp 12 S \left| J_1 \gamma_1 \left( \mathbf{k} \right) + J_3 \gamma_3 \left( \mathbf{k}\right) \right|,	\label{eq:magnonenergy}
\end{align}
with $-$ ($+$) for $\epsilon_1$ ($\epsilon_2$), and eigenvectors
\begin{align}
\psi_{1/2}	&=	\frac{1}{\sqrt{2}} \left( \mp f\left(\mathbf{k}\right),+1\right)^T,	\label{eq:psi}
\end{align}
where
\begin{align}
f\left(\mathbf{k}\right)	&=	\frac{J_1 \gamma_1 \left(\mathbf{k}\right) + J_3 \gamma_3 \left(\mathbf{k}\right)}{\left| J_1 \gamma_1 \left(\mathbf{k}\right) + J_3 \gamma_3 \left(\mathbf{k}\right)\right|} \equiv \frac{g\left( \mathbf{k}\right)}{|g\left( \mathbf{k}\right)|}	\label{eq:functionf},
\end{align}
i.e. the states have the same structure as in Eq.~\eqref{eq:general:psi}. 
The gap $2\Delta\epsilon (\mathbf{k}) = \epsilon_2 (\mathbf{k}) - \epsilon_1 (\mathbf{k}) = 24 S \left| J_1 \gamma_1 \left( \mathbf{k} \right) + J_3 \gamma_3 \left( \mathbf{k}\right) \right| = 24S |g(\mathbf{k})|$ only depends on the inter-sublattice interactions $J_{1,3}$. Non-accidental degeneracies occur when $\gamma_1 (\mathbf{k})=\gamma_3\left(\mathbf{k}\right)=0$. The structure of Eqs.~\eqref{eq:gamma1}, \eqref{eq:gamma3} (or Eqs.~\eqref{eq:gamma1periodic}, \eqref{eq:gamma3periodic}) is such that this occurs either when the first factor vanishes, $\left(1+e^{i c k_z}\right)=0,$ or when the second factors vanish. At $k_z=\pm \pi/c$ ($\ell=\pm 1/2$), $\left(1+e^{i c k_z}\right)=0\, \forall k_x, k_y,$ which produces the nodal planes. The second factors vanish at the $K$, $K'$ points (which are related by a $C_6$ rotation), and along paths $K\rightarrow H\rightarrow K$, $K'\rightarrow H'\rightarrow K'$ at finite $k_z$, giving rise to the nodal lines.


In Section~\ref{sec:symtop} we argued that the nodal plane is protected by a combination of effective time reversal and screw symmetries. The time reversal symmetry can be seen explicitly in the $J_1 - J_4$ model from the identity $\gamma_n \left( -\mathbf{k}\right) = \gamma_n^\star \left( \mathbf{k}\right)$ and the fact that the linear spin wave Hamiltonian depends exclusively on these functions. The screw symmetry places constraints Eqs.~(\ref{eq:2foldscrewA}) and (\ref{eq:2foldscrewB}) on the Hamiltonian and it is straightforward to check that both are satisfied by $C_d(\mathbf{k})$ and $C_0(\mathbf{k})$ respectively. The latter is true because $k_z$ appears only through $(1+e^{ick_z})$, in the $\eta=0$ convention, which equals $e^{ick_z}(1+e^{-ick_z})$.

As mentioned in the main text, the nodal lines can be classified in terms of a closed-path Berry phase,
\begin{align}
\gamma_m \left[ \mathcal{C} \right] = \oint_\mathcal{C} \mathrm{d}\mathbf{k} \cdot \mathcal{A}_m \left( \mathbf{k}\right),
\end{align}
where $\mathcal{C}$ is a closed contour, $\mathcal{A}$ is the Berry connection,
\begin{align}
\mathcal{A}_m	&=	i \left\langle u_m \left( \mathbf{k}\right) \middle| \nabla_\mathbf{k} \middle| u_m \left( \mathbf{k}\right) \right\rangle,
\end{align}
and $\left| u_m \left( \mathbf{k}\right)\right\rangle$ is an eigenstate of $h(\mathbf{k})$. Using Eq.~\eqref{eq:psi}, 
\begin{equation}
\mathcal{A}	=	\mathcal{A}_m = \frac{i}{2} f^\star(\mathbf{k}) \nabla_\mathbf{k} f(\mathbf{k}),	\quad \forall m,   \label{eq:connection}
\end{equation}
from which it is clear that the topological properties are related to the intersublattice couplings $J_{1,3}$, and independent of $J_{2,4}$. (Thus the $J_1$, $J_2$, $J_3$ model considered in the main text has identical topology to the model here.)
To obtain $\gamma_m$, it is convenient to shift the $\mathbf{k}$-space origin to e.g. $K$, using coordinates $(k_x',k_y',k_z')$ and then introduce cylindrical coordinates,
\begin{align}
k_x'	&=	\rho \cos \varphi,\quad k_y'=\rho \sin \varphi,	\quad k_z'=k_z.
\end{align}
such that $\rho$ describes the radius of a circular loop about the nodal line, and $\varphi$ the angle along the loop. Then direct evaluation of Eq.~\eqref{eq:connection} (here performed using Mathematica at various $k_z$ values) yields $\gamma=\pm \pi$ at $K$ and $K'$, see Table~\ref{table:Berry}.

\begin{table}
	\caption{\label{table:Berry}Berry phase calculated at loops surrounding different lines in momentum space. The leftmost column indicates a line segment, the $\mathbf{k}$-space coordinates of which are given in two coordinate systems (columns two and three).}
	\begin{center}
		\begin{tabular}{ |c|c|c|c| } 
			\hline
			Point   & $(k_x,k_y,k_z)$ & $(hk\ell)$ & $\gamma_1 = \gamma_2$\\\hline
			$\Gamma-A-\Gamma$& $\left( 0,0,k_z\right)$		& $\left( 0,0,\ell 	\right)$	& 0\\
			$K-H-K$     &$\left( \frac{2\pi}{3\alpha}, \frac{2\pi}{\sqrt{3}\alpha},k_z \right)$ & $\left( \frac{1}{3}, \frac{1}{3}, \ell\right)$  & $\pi$\\ 
			$K-H-K$     &$\left( -\frac{4\pi}{3\alpha}, 0,k_z \right)$ & $\left( -\frac{2}{3}, \frac{1}{3}, \ell\right)$ &	$\pi$\\ 
			$K-H-K$     &$\left( \frac{2\pi}{3\alpha}, -\frac{2\pi}{\sqrt{3}\alpha},k_z \right)$ & $\left( \frac{1}{3}, -\frac{2}{3}, \ell\right)$	& $\pi$\\
			$K'-H'-K'$     &$\left( -\frac{2\pi}{3\alpha}, -\frac{2\pi}{\sqrt{3}\alpha},k_z \right)$ & $\left( -\frac{1}{3}, -\frac{1}{3}, \ell\right)$ & $-\pi$\\
			$K'-H'-K'$     &$\left( \frac{4\pi}{3\alpha}, 0,k_z \right)$ & $\left( \frac{2}{3}, -\frac{1}{3}, \ell\right)$ & $-\pi$\\ 
			$K'-H'-K'$     &$\left( -\frac{2\pi}{3\alpha}, \frac{2\pi}{\sqrt{3}\alpha},k_z \right)$ & $\left( -\frac{1}{3}, \frac{2}{3}, \ell\right)$	& $-\pi$\\
			\hline
		\end{tabular}
	\end{center}
\end{table}

\begin{figure}
	\centering\includegraphics[width=0.48\textwidth]{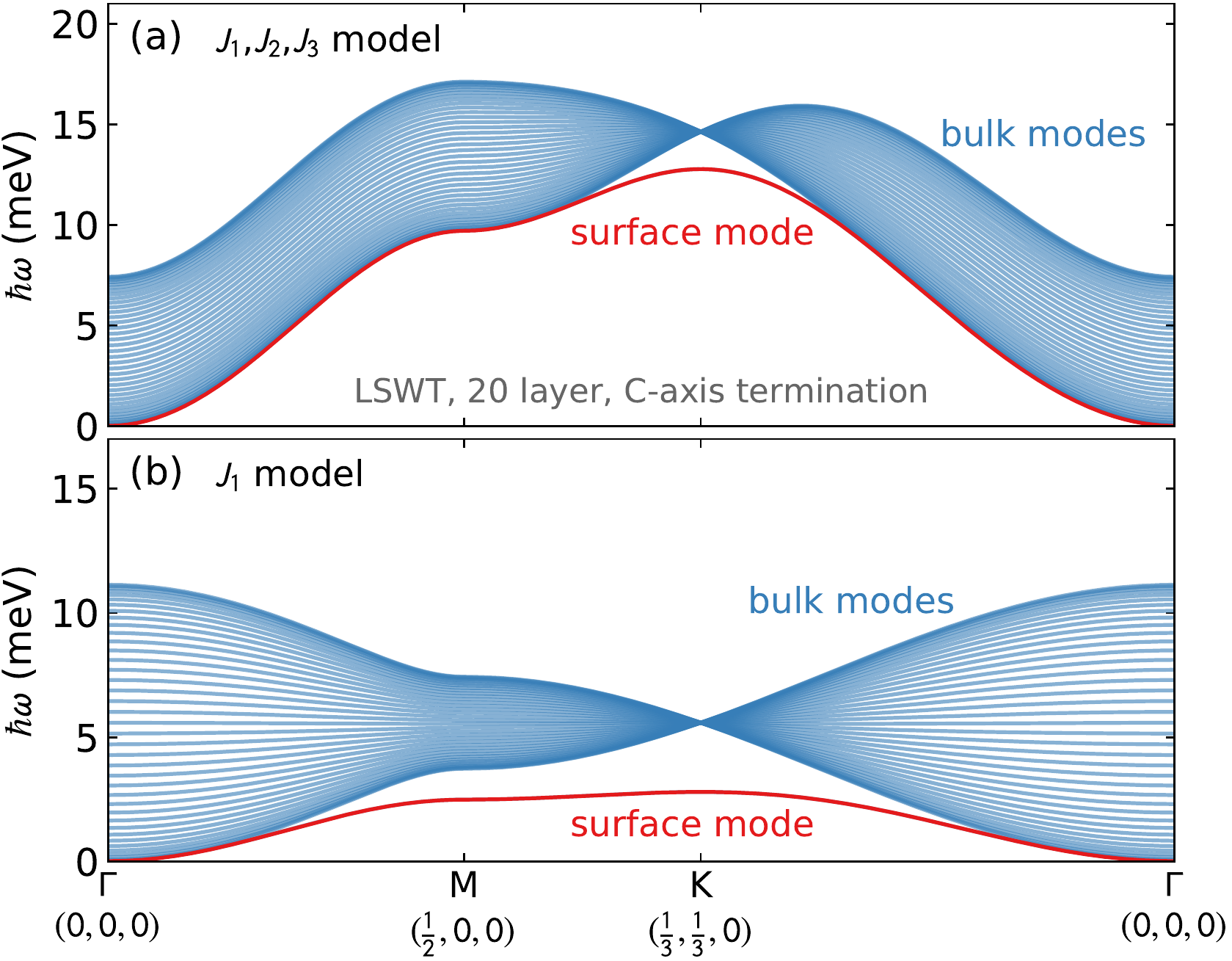}
	\caption{Surface magnons for a simplified Linear spin wave theory (LSWT) models, showing the topological band crossings and surface modes are independent of the particular model details. (a) Surface magnon calculated for a 20-layer Gd slab using the three nearest neighbor fitted exchange interactions. (b) Surface magnon calculated for the nearest neighbor exchange only. Note that in both cases, the bulk modes linearly cross at at $K$, while the surface magnon mode lies outside this continuum, indicating that this surface magnon is a property of the lattice symmetry rather than the details of the Hamiltonian.}
	\label{flo:BulkSurface3}
\end{figure}
As noted in the main text, the nodal line gives rise to a clear topological surface mode. The degree to which it separates from the bulk modes does depend on the specific exchanges included in the Hamiltonian. This is illustrated by the results for a pure $J_1$ model and the $J_1-J_2-J_3$ model shown in Fig.~\ref{flo:BulkSurface3}, using the same slab geometry supercell as for the full fitted model.

\end{document}